\documentclass{article}
\usepackage{arxiv}
\usepackage[T1]{fontenc}    
\usepackage{hyperref}       
\usepackage{url}            
\usepackage{booktabs}       
\usepackage{amsmath,amssymb,amsfonts}
\usepackage{nicefrac}       
\usepackage{microtype}      
\usepackage{lipsum}
\usepackage{graphicx}
\usepackage{multirow}
\usepackage{color}
\usepackage[most]{tcolorbox}
\usepackage{array}
\definecolor{lightgray}{gray}{0.95}
\usepackage[most]{tcolorbox}
\usepackage{tabularx}
\usepackage{enumitem}
\graphicspath{ {./images/} }
\usepackage{authblk}
\usepackage{amsmath}
\usepackage{cite}
\usepackage{amsmath,amssymb,amsfonts}
\usepackage{bbm}
\usepackage{algorithmic}
\usepackage{graphicx}
\usepackage{textcomp}
\usepackage{xcolor}
\usepackage{colortbl}
\definecolor{lightgray}{gray}{0.95}
\usepackage{booktabs}
\usepackage{makecell}
\usepackage{multirow}
\usepackage{pifont} 
\usepackage{booktabs} 
\usepackage{subfigure}
\usepackage[most]{tcolorbox}
\def\BibTeX{{\rm B\kern-.05em{\sc i\kern-.025em b}\kern-.08em
    T\kern-.1667em\lower.7ex\hbox{E}\kern-.125emX}}

\title{Generating Is Believing: Membership Inference Attacks against Retrieval-Augmented Generation}

\author[1]{Yuying Li}
\author[2,*]{Gaoyang Liu}
\author[2]{Chen Wang}
\author[1]{Yang Yang}

\affil[1]{Key Laboratory of Intelligent Sensing System and Security (Ministry of Education) \newline School of Artificial Intelligence, Hubei University, Wuhan, China}
\affil[2]{Huazhong University of Science and Technology, Wuhan, China \newline yuyingli@stu.hubu.edu.cn, yangyang@hubu.edu.cn, \{husterlgy, cwangwhu\}@gmail.com}

\begin{document}
\maketitle 
\begin{abstract}
Retrieval-Augmented Generation (RAG) is a state-of-the-art technique that mitigates issues such as hallucinations and knowledge staleness in Large Language Models (LLMs) by retrieving relevant knowledge from an external database to assist in content generation. Existing research has demonstrated potential privacy risks associated with the LLMs of RAG. However, the privacy risks posed by the integration of an external database, which often contains sensitive data such as medical records or personal identities, have remained largely unexplored.  
In this paper, we aim to bridge this gap by focusing on membership privacy of RAG's external database, with the aim of determining whether a given sample is part of the RAG's database. Our basic idea is that if a sample is in the external database, it will exhibit a high degree of semantic similarity to the text generated by the RAG system. We present S$^2$MIA, a \underline{M}embership \underline{I}nference \underline{A}ttack that utilizes the \underline{S}emantic \underline{S}imilarity between a given sample and the content generated by the RAG system. With our proposed S$^2$MIA, we demonstrate the potential to breach the membership privacy of the RAG database.  
Extensive experiment results demonstrate that S$^2$MIA can achieve a strong inference performance compared with five existing MIAs, and is able to escape from the protection of three representative defenses. 
\end{abstract}

\section{Introduction}
Retrieval-Augmented Generation (RAG) is an advanced natural language processing technique that enhances the content generation capabilities of Large Language Models (LLMs) by integrating information retrieved from an external database. This technique addresses several limitations of LLMs, such as hallucinations, knowledge staleness, and knowledge gaps in particular domains (e.g., medical domain)~\cite{shi2024replug, shang2024ontofact, zhao2024ltgc, asai2023self, deng2023regavae,wang2024retrieval}. 

Despite these advancements in LLMs, RAG systems face serious security and privacy challenges. Existing works have demonstrated that RAG systems are vulnerable to various attacks, such as prompt prompt injection~\cite{zou2024poisonedrag,zeng2024good}, jailbreaks~\cite{deng2024pandora}, and prompt perturbation~\cite{hu2024prompt}, which can lead the RAG system to generate incorrect or harmful content. However, these studies primarily focus on misleading the LLM component of RAG systems. The security of RAG's external databases, which often contain sensitive data such as medical records or personal identities, remains largely unexplored~\cite{anderson2024my,zeng2024good}.

In this work, we focus on a fundamental attack known as Membership Inference Attacks (MIAs)~\cite{ko2023practical,meeus2024did,maini2024llm} to evaluate and measure the privacy and security of the RAG’s external database. The goal of MIAs against RAG is to determine whether a given sample (i.e., the target sample) is part of the external database of the given RAG system. Successful MIAs can lead to severe privacy risks for the data providers of RAG systems. For example, if a target sample in a medical domain RAG's database is identified, it may reveal the disease history of the data provider.

However, existing MIAs are typically tailored for traditional classification or generative machine learning models. These approaches rely on the model's tendency to overfit its training samples, which results in distinguishable predictions for the training samples. In contrast, RAG systems do not train the LLM on their external databases and lack the overfitting characteristic on the samples within the external databases. Consequently, existing MIA methods cannot be directly applied to RAG systems. 
Recently, Anderson et al.~\cite{anderson2024my} provide an MIA against RAG, which directly prompts the RAG system to answer whether a specific sample exists in the database.  
This attack relies on the LLM's output within the RAG system, and since LLMs responses are generated through sample techniques (e.g., top-k~\cite{minaee2024large} and top-p~\cite{zhou2024balancing} sampling), the target sample may not consistently align with the output, making it difficult for performing MIAs. 

\begin{figure}[t]
\includegraphics[width=0.7\linewidth]{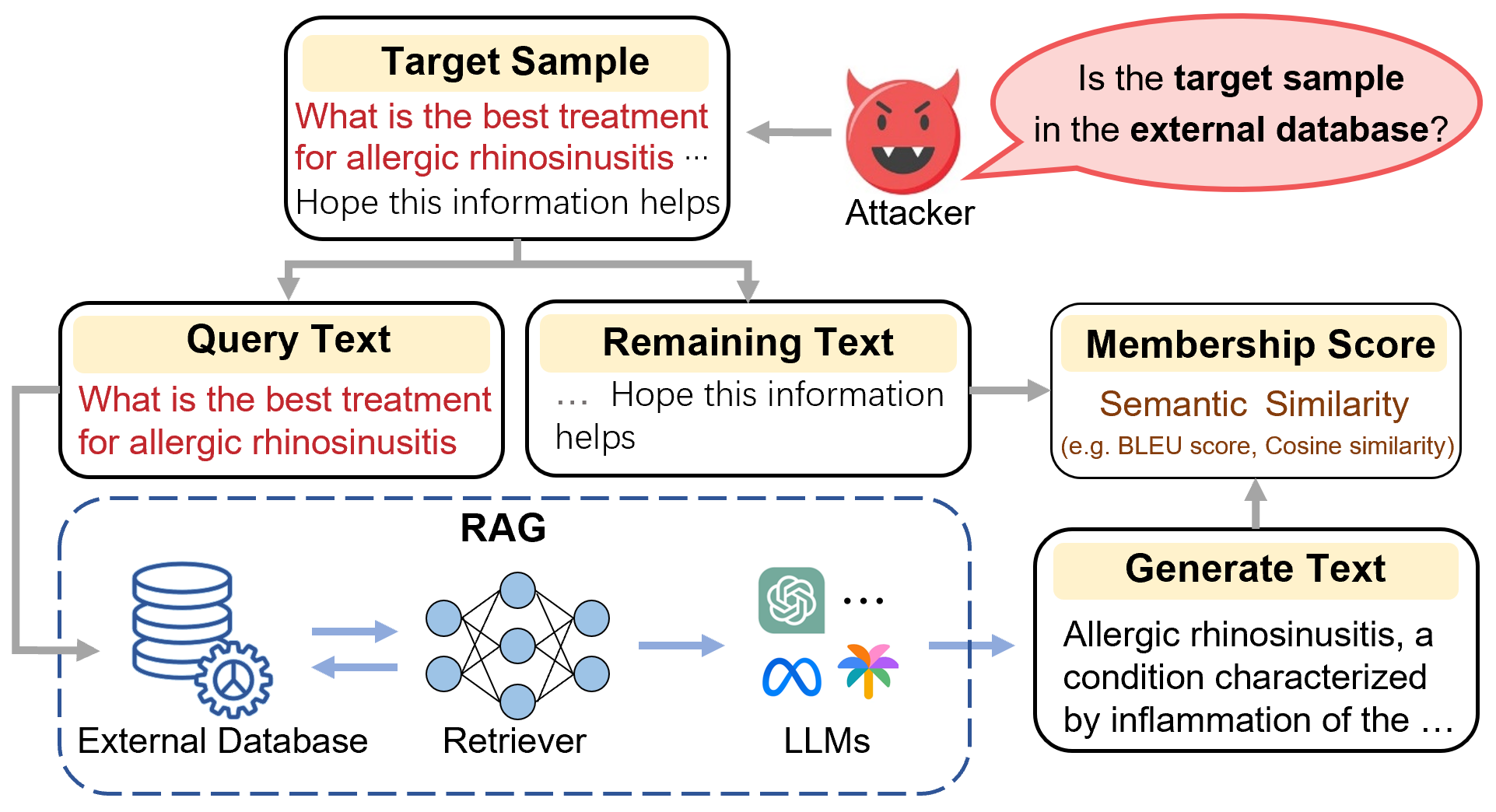}
\centering
\caption{Framework of S$^2$MIA. Given a target sample, we divide it into the query text and the remaining text. We input the query text into a given RAG system, which retrieves the multiple samples most similar to the query test from the external database. These retrieved texts serve as the context for the LLM to generate responses. Consequently, if the target sample is present in the RAG's external database, the generated text should exhibit high semantic similarity and low perplexity to the remaining text. } 
    \label{fig1}
\end{figure}

In this paper, we propose a novel MIA against RAG systems, named S$^2$MIA, which leverage the \underline{S}emantic \underline{S}imilarity between the target sample and the generated content to perform \underline{MIA}, as illustrated in Figure~\ref{fig1}. 
The basic idea of our MIA is that if a sample is in the RAG system's external database, the generated content will be similar to this sample when querying the given RAG system with this sample.
This occurs because the RAG system would first retrieve the samples from its external database that are most similar to the input~\cite{morris2023text,shi2023large,radit}, and then integrate the retrieved sample to generate the output. If a target sample is within the database, this process enables the RAG system to generate an output that exhibits a higher degree of similarity to the target sample.
By comparing the similarity between the target sample and the content generated by the RAG system, we can perform MIA and breach the membership privacy against the target sample. 

Although our concept is straightforward, implementing S$^2$MIA faces a significant challenge. RAG systems utilize both retrieve and target samples to query LLMs for output generation. However, the sampling techniques of LLMs prevent the output content from perfectly matching the target sample. Consequently, directly comparing the similarity between the generated text and the target sample poses a challenge for effectively performing MIA. 
In the RAG system, if the target sample exists in an external database, 
then the retriever inside the RAG will be used to retrieve multiple samples similar to the target sample, including the target sample itself, for querying LLMs to generate the output. Although these generated outputs may differ in the explicit representation, they still exhibit semantic similarity with the target sample. 
Therefore, to tackle this issue, we can first calculate the semantic similarity between the target sample and the generated output, and then utilize this similarity as the membership feature to conduct MIA against RAG systems. 


This paper makes the following contributions:
\begin{itemize}
    \item We present S$^2$MIA, a novel MIA against RAG systems, which addresses the significant gap in understanding the privacy vulnerabilities of RAG's external databases. 
    \item We find the semantic similarity difference between samples that are part of the RAG system's external database and those that are not concerning their corresponding generated contents, and show for the first time that the semantic similarity may leak the membership privacy and can be leveraged to perform MIA against RAG systems. 
    \item We evaluate the performance of S$^2$MIA on five RAG systems, and results demonstrate that S$^2$MIA can achieve a strong performance compared with four existing MIAs and successfully bypass three defenses. 
\end{itemize}

\section{Preliminary}
\subsection{Retrieval-Augmented Generation}
RAG systems~\cite{fan2024survey,mallen2023not} typically consist of three components: an external database, a retriever, and an LLM. The external database contains a set of text samples collected from specific domains, which is denoted as $D = \{d_1, d_2, ..., d_n\}$, where $d_i$ is the text of the $i$-th sample. 
To generate text content from an input query,  RAG systems generally operate in two main stages: retrieving data and generating outputs.  
Upon receiving a query $Q$,  RAG system calculates the similarity between $Q$ and each samples’s in $D$:
\begin{equation}
S = Sim(Q, d_i), 
\end{equation}
where $S$ is the sample similarity, and $Sim$ can be any similarity metric, such as cosine similarity. 
Based on $S$, the retriever finds the most relevant texts, forming a set $D_s$:
\begin{equation}
D_s = \text{R}(Q, D, S),
\end{equation}
where $R$ is the retriever of RAG. Finally, the RAG system integrates the query $Q$ with $D_s$ to formulate a prompt for the LLM $M$, which then generates the output.

\subsection{Threat Model}
\noindent \textbf{Attacker’s Goal: }Given a target sample $x_t$ and a RAG system, the attacker’s goal is to infer whether $x_t$ is in the external database $D$ or not: 
\begin{equation} 
\mathcal{A}(x_t \mid D, R, M) \rightarrow \textbf{Member} / \textbf{Non-Member},
\end{equation} 
where $\mathcal{A}$ represents the MIA method, and \textbf{Member} (resp. \textbf{Non-Member}) indicates that $x_t$ is (resp. not) a part of $D$. 

\noindent \textbf{Attacker’s Knowledge: }The attacker does not have access to the LLM parameters, the configuration or operation details of the Retriever, nor any sample within the external database. They only have the distribution of the external database. 

\noindent \textbf{Attacker’s Capability: }As in previous works~\cite{shi2024replug,panda2024privacy}, the attacker in our paper can query RAG systems and obtain the output text along with the prediction probabilities.

\section{Method}
To evaluate the membership privacy of a target RAG's external database, we have developed an MIA named S$^2$MIA. 
We utilize the semantic similarity between the outputs generated by RAG and the target sample, and then contrast this similarity with the differences observed between member and non-member samples in RAG's external database to perform MIAs.
Overall, S$^2$MIA mainly consists of the following two parts (c.f. Fig.~\ref{fig1}).

\subsection{Membership Score Generation} \label{score}
Given a target sample $x_t$, we first divide it into two parts: the query text $x_t^q$ and the remaining text $x_t^r$, formatted as $x_t = {x_t^q \oplus x_t^r}$. Then we query the target RAG system with $x_t^q$ with the following prompt: 
\begin{tcolorbox}[
    colframe=black!75,          
    colback=gray!10,            
    coltitle=black,             
    sharp corners,              
    width=\linewidth,           
    boxrule=0.5mm,              
    fontupper=\fontsize{8pt}{8pt}\selectfont, 
    boxsep=1mm,                 
    left=1mm,                   
    right=1mm,                  
    top=0.6mm,                    
    bottom=0.6mm                  
]
Given the [Query]: \texttt{"query"}.\\
Do not include any introductory or explanatory text, use the following format for output: 
\{[Response]: 'Provide a concise response directly addressing the [Query] by using the most relevant and matching text in the prompt.'\} 
\end{tcolorbox}

\begin{table*}[t]
\caption{ROC AUC and PR AUC of S$^2$MIA and five comparison methods (The highest AUC values for each dataset are highlighted in bold). }\label{tab1}
\centering
\renewcommand{\arraystretch}{0.9}
\fontsize{8pt}{8pt}\selectfont
\arrayrulecolor[gray]{0.7}
\resizebox{\textwidth}{!}{%
\begin{tabular}{c|ccccc|ccccc}
\Xhline{3\arrayrulewidth}
 {\raisebox{-0.15em}{Dataset}} & \multicolumn{5}{c|}{\raisebox{-0.15em}{Natural Question}} & \multicolumn{5}{c}{\raisebox{-0.15em}{Trivia-QA}}\\
\hline
\arrayrulecolor{black}
\raisebox{-0.15em}{Model} & \raisebox{-0.15em}{Llama2-7b} & \raisebox{-0.15em}{Llama2-13b} & \raisebox{-0.15em}{Vicuna} & \raisebox{-0.15em}{Alpaca} & \raisebox{-0.15em}{GPT-3.5} 
      & \raisebox{-0.15em}{Llama2-7b} & \raisebox{-0.15em}{Llama2-13b} & \raisebox{-0.15em}{Vicuna} & \raisebox{-0.15em}{Alpaca} & \raisebox{-0.15em}{GPT-3.5} \\
\hline \hline
 {\raisebox{-0.15em}{Metric}} & \multicolumn{10}{c}{\raisebox{-0.15em}{ROC AUC}}\\
 \arrayrulecolor[gray]{0.7}\hline
\arrayrulecolor{black}
\raisebox{-0.15em}{Loss Attack} & \raisebox{-0.15em}{0.520} & \raisebox{-0.15em}{0.533} & \raisebox{-0.15em}{0.508} & \raisebox{-0.15em}{0.505} & \raisebox{-0.15em}{0.517} & \raisebox{-0.15em}{0.516} & \raisebox{-0.15em}{0.531} & \raisebox{-0.15em}{0.536} & \raisebox{-0.15em}{0.508} & \raisebox{-0.15em}{0.518} \\
\raisebox{-0.15em}{Zlib Entropy Attack} & \raisebox{-0.15em}{0.537} & \raisebox{-0.15em}{0.518} & \raisebox{-0.15em}{0.515} & \raisebox{-0.15em}{0.509} & \raisebox{-0.15em}{0.502} & \raisebox{-0.15em}{0.519} & \raisebox{-0.15em}{0.509} & \raisebox{-0.15em}{0.513} & \raisebox{-0.15em}{0.563} & \raisebox{-0.15em}{0.512} \\
\raisebox{-0.15em}{Min k\% Prob Attack} & \raisebox{-0.15em}{0.576} & \raisebox{-0.15em}{0.519} & \raisebox{-0.15em}{0.503} & \raisebox{-0.15em}{0.549} & \raisebox{-0.15em}{0.528} & \raisebox{-0.15em}{0.535} & \raisebox{-0.15em}{0.523} & \raisebox{-0.15em}{0.514} & \raisebox{-0.15em}{0.485} & \raisebox{-0.15em}{0.505} \\
\raisebox{-0.15em}{Neighborhood Attack} & \raisebox{-0.15em}{0.511} & \raisebox{-0.15em}{0.504} & \raisebox{-0.15em}{0.518} & \raisebox{-0.15em}{0.515} & \raisebox{-0.15em}{0.502} & \raisebox{-0.15em}{0.521} & \raisebox{-0.15em}{0.538} & \raisebox{-0.15em}{0.510} & \raisebox{-0.15em}{0.516} & \raisebox{-0.15em}{0.521} \\
\raisebox{-0.15em}{RAG-MIA} & \raisebox{-0.15em}{0.830} & \raisebox{-0.15em}{0.815} & \raisebox{-0.15em}{0.854} & \raisebox{-0.15em}{0.826} & \raisebox{-0.15em}{0.806} & \raisebox{-0.15em}{0.821} & \raisebox{-0.15em}{0.781} & \raisebox{-0.15em}{0.780} & \raisebox{-0.15em}{0.795} & \raisebox{-0.15em}{0.761} \\
\rowcolor{lightgray}
\raisebox{-0.15em}{S$^2$MIA-T} & \raisebox{-0.15em}{\textbf{0.892}} & \raisebox{-0.15em}{0.877} & \raisebox{-0.15em}{\textbf{0.874}} & \raisebox{-0.15em}{\textbf{0.881}} & \raisebox{-0.15em}{\textbf{0.884}} & \raisebox{-0.15em}{\textbf{0.885}} & \raisebox{-0.15em}{0.765} & \raisebox{-0.15em}{0.772} & \raisebox{-0.15em}{\textbf{0.869}} & \raisebox{-0.15em}{\textbf{0.856}} \\
\rowcolor{lightgray}
\raisebox{-0.15em}{S$^2$MIA-M} & \raisebox{-0.15em}{0.878} & \raisebox{-0.15em}{\textbf{0.893}} & \raisebox{-0.15em}{0.867} & \raisebox{-0.15em}{0.863} & \raisebox{-0.15em}{0.881} & \raisebox{-0.15em}{0.871} & \raisebox{-0.15em}{\textbf{0.794}} & \raisebox{-0.15em}{\textbf{0.801}} & \raisebox{-0.15em}{0.798} & \raisebox{-0.15em}{0.837} \\

\hline \hline
 {\raisebox{-0.15em}{Metric}} & \multicolumn{10}{c}{\raisebox{-0.15em}{PR AUC}}\\
 \arrayrulecolor[gray]{0.7}\hline
\arrayrulecolor{black}
\raisebox{-0.15em}{Loss Attack} & \raisebox{-0.15em}{0.537} & \raisebox{-0.15em}{0.518} & \raisebox{-0.15em}{0.503} & \raisebox{-0.15em}{0.521} & \raisebox{-0.15em}{0.507} & \raisebox{-0.15em}{0.504} & \raisebox{-0.15em}{0.525} & \raisebox{-0.15em}{0.511} & \raisebox{-0.15em}{0.504} & \raisebox{-0.15em}{0.503} \\
\raisebox{-0.15em}{Zlib Entropy Attack} & \raisebox{-0.15em}{0.528} & \raisebox{-0.15em}{0.509} & \raisebox{-0.15em}{0.515} & \raisebox{-0.15em}{0.507} & \raisebox{-0.15em}{0.517} & \raisebox{-0.15em}{0.516} & \raisebox{-0.15em}{0.501} & \raisebox{-0.15em}{0.522} & \raisebox{-0.15em}{0.554} & \raisebox{-0.15em}{0.505} \\
\raisebox{-0.15em}{Min k\% Prob Attack} & \raisebox{-0.15em}{0.556} & \raisebox{-0.15em}{0.528} & \raisebox{-0.15em}{0.507} & \raisebox{-0.15em}{0.539} & \raisebox{-0.15em}{0.537} & \raisebox{-0.15em}{0.518} & \raisebox{-0.15em}{0.529} & \raisebox{-0.15em}{0.507} & \raisebox{-0.15em}{0.505} & \raisebox{-0.15em}{0.503} \\
\raisebox{-0.15em}{Neighborhood Attack} & \raisebox{-0.15em}{0.511} & \raisebox{-0.15em}{0.526} & \raisebox{-0.15em}{0.514} & \raisebox{-0.15em}{0.514} & \raisebox{-0.15em}{0.507} & \raisebox{-0.15em}{0.511} & \raisebox{-0.15em}{0.527} & \raisebox{-0.15em}{0.501} & \raisebox{-0.15em}{0.526} & \raisebox{-0.15em}{0.513} \\
\raisebox{-0.15em}{RAG-MIA} & \raisebox{-0.15em}{0.819} & \raisebox{-0.15em}{0.827} & \raisebox{-0.15em}{0.845} & \raisebox{-0.15em}{0.819} & \raisebox{-0.15em}{0.824} & \raisebox{-0.15em}{0.851} & \raisebox{-0.15em}{0.796} & \raisebox{-0.15em}{0.792} & \raisebox{-0.15em}{0.815} & \raisebox{-0.15em}{0.791} \\
\rowcolor{lightgray}
\raisebox{-0.15em}{S$^2$MIA-T} & \raisebox{-0.15em}{0.872} & \raisebox{-0.15em}{0.884} & \raisebox{-0.15em}{\textbf{0.864}} & \raisebox{-0.15em}{0.851} & \raisebox{-0.15em}{\textbf{0.893}} & \raisebox{-0.15em}{\textbf{0.875}} & \raisebox{-0.15em}{0.791} & \raisebox{-0.15em}{0.782} & \raisebox{-0.15em}{\textbf{0.869}} & \raisebox{-0.15em}{\textbf{0.878}} \\
\rowcolor{lightgray}
\raisebox{-0.15em}{S$^2$MIA-M} & \raisebox{-0.15em}{\textbf{0.882}} & \raisebox{-0.15em}{\textbf{0.903}} & \raisebox{-0.15em}{0.854} & \raisebox{-0.15em}{\textbf{0.872}} & \raisebox{-0.15em}{0.869} & \raisebox{-0.15em}{0.875} & \raisebox{-0.15em}{\textbf{0.803}} & \raisebox{-0.15em}{\textbf{0.799}} & \raisebox{-0.15em}{0.827} & \raisebox{-0.15em}{0.856} \\
\Xhline{3\arrayrulewidth}
\end{tabular}
}
\vspace{-5pt}
\end{table*}

In the operation of RAG systems, the retriever selects multiple samples $D_s$ from the external database $D$ and concatenates $D_s$ with $x_t^q$ the query text to form a prompt for the LLM. If $x_t$ is in $D$, it is likely to be included in $D_s$ and influence the LLM's output. 
However, even with a prompt designed to highlight the most relevant sample, the LLM's generated text $\bar{x}^g$ will not perfectly match the target sample, due to the inherent sampling techniques~\cite{minaee2024large,zhou2024balancing} that introduce randomness into the generation process.
Therefore, directly comparing the similarity between 
$x_t$ and $\bar{x}^g$ will hinder the inference process of MIAs.

To address this issue, we focus on utilizing the semantic similarity, which is quantified by the BLEU score~\cite{devlin2018bert} between $x_t$ and $\bar{x}^g$, denoted as $S_{sem}$:
\begin{equation}
S_{sem} = BLEU(x_t, \bar{x}^g). 
\end{equation}

In order to enhance the performance of S$^2$MIA, we also incorporate the perplexity of the generated text of RAG. 

Since the RAG system generates responses based on the target sample $x_t$ and its similar samples $D_s$, if $x_t$ is in the external database $D$, the generated $\bar{x}^g$'s average uncertainty in predicting each word should be low. This implies that member samples tend to exhibit low perplexity. 
Then we calculate the generation $\bar{x}^g$'s perplexity, demoted as $PPL_{gen}$, and use $S_{sem}$ and $PPL_{gen}$ as membership score of $x_t$. 

\subsection{Membership Inference}
Given the membership score of the target sample, we propose two methods, S$^2$MIA-T, and S$^2$MIA-M, for conducting the subsequent membership inference. 

\noindent \textbf{Threshold-based Attack (S$^2$MIA-T). }Assuming the attacker has knowledge of the data distribution of the external database $D$, we can extract multiple samples from this distribution, which are labeled as \textbf{members}. Subsequently, we sample \textbf{non-members} from other distributions. We combine these member and non-member samples as a reference dataset $D_r$ to determine the optimal threshold for the membership score. 

We utilize a greedy search method to determine the optimal thresholds, $\theta_{sem}^*$ for the semantic similarity and $\theta_{gen}^*$ for the generation perplexity. In this process, we adjust each threshold individually while holding the other constant. We classify all samples of $D_r$ using the current combination of thresholds and select the combination that results in the best attack performance.
Then the target  sample is classified as a member if $S_{sem} \geq \theta_{sem}^*$ and $PPL_{gen} \leq \theta_{gen}^*$; otherwise, it is classified as a non-member. 

\noindent \textbf{Model-based Attack (S$^2$MIA-M). }In addition, we utilize supervised machine learning models to conduct MIA on the membership scores $S_{sem}$ and $PPL_{gen}$. 
For each sample in $D_r$, we calculate the semantic similarity and the generation perplexity. These are combined to form a membership feature vector, which is then labeled with the corresponding membership property. We then train an attack classification model, and infer the target sample's membership property $f_{MIA}(S_{sem,x_t}, PPL_{gen,x_t}) $, where $S_{sem,x_t}$ and $PPL_{gen,x_t}$ is the target sample's membership features. 
It should be noted that various supervised classification methods, including neural networks and XGBoost, can be employed in our inference.

\section{Experiment and result}
\subsection{Experiment Settings}

\noindent \textbf{Dataset. }We use two datasets commonly employed in RAG attack research: Natural Questions~\cite{NQ} and Trivia-QA~\cite{joshi2017triviaqa}, in accordance with existing works~\cite{zou2024poisonedrag,zeng2024mitigatingrag,kim2024re}.

\noindent \textbf{RAG. }We use the standard RAG framework~\cite{guu2020retrieval} in our experiments. We respectively use LLaMA-2-7b-chat-hf~\cite{touvron2023llama}, LLaMA-2-13b-chat-hf~\cite{touvron2023llama}, Vicuna~\cite{chiang2023vicuna}, Alpaca~\cite{alpaca}, and GPT-3.5-turbo~\cite{gpt3.5} as the LLM component of RAG systems. We use Contreiver~\cite{izacard2021contriever} and DPR~\cite{karpukhin-etal-2020-dpr} as the retrievers, setting them to retrieve the 5 most similar samples from the external database. For other parameters, we use the default settings as specified in ~\cite{guu2020retrieval}.

\noindent \textbf{Attack Settings. }For each dataset, we randomly select 80\% of the samples to form the RAG system's external database (i.e., member samples), while the remaining 20\% are designated as non-member samples.
For each dataset, we randomly select $1,000$ member samples and $1,000$ non-member samples to construct the reference dataset. Subsequently, we apply the same methodology to select an additional $2,000$ samples, which are used to evaluate the performance of S$^2$MIA.
Specifically, each sample's question part (resp. answer part) is treated as the query text (resp. remaining text). 

\noindent \textbf{Evaluation Metrics. }Following previous MIAs~\cite{duan2024membership,anderson2024my,duan2023diffusion}, we utilize ROC AUC and PR AUC as the metrics. 

\noindent \textbf{Baselines. }
Our comparison baselines include:
\begin{itemize}
    \item  \textbf{Loss Attack}~\cite{yeom2018loss}: this attack performs MIAs by using the model loss on the target samples.
    \item  \textbf{Zlib Entropy Attack}~\cite{carlini2021zlib}: this attack performs MIAs using the zlib entropy of the target samples. 
    \item  \textbf{Neighborhood Attack}~\cite{mattern2023neighbor}: this attack generates neighbor samples around the target sample and then compares the losses of neighbor samples to perform MIAs.
    \item  \textbf{Min-k\% Prob Attack}~\cite{shi2023mink}: this attack uses the k\% tokens with minimum probabilities to execute MIAs. 
    \item \textbf{RAG-MIA}~\cite{anderson2024my}: this attack directly queries the target RAG system to infer whether a target sample is part of its external database.
\end{itemize}

\begin{table}[t]
\caption{Impact of the similarity metric on S$^2$MIA-T and S$^2$MIA-M in Natural Question dataset}\label{tab2}
\vspace{-5pt}
\centering
\setlength{\tabcolsep}{2pt} 
\arrayrulecolor{black}
\renewcommand{\arraystretch}{0.8}
\begin{tabular}{cccccc}
\Xhline{3\arrayrulewidth}
\multirow{2}{*}{\raisebox{-0.3em}{Similarity}} & \multicolumn{2}{c}{\raisebox{-0.2em}{S$^2$MIA-T}} & \multicolumn{2}{c}{\raisebox{-0.2em}{S$^2$MIA-M}} \\
\cmidrule(r){2-3} \cmidrule(r){4-5}
 & ROC AUC & PR AUC & ROC AUC & PR AUC  \\  
\hline
\raisebox{-0.15em}{BLEU(w/ Perplexity)}  & \raisebox{-0.15em}{0.892} & \raisebox{-0.15em}{0.872} & \raisebox{-0.15em}{0.878} & \raisebox{-0.15em}{0.882} \\
\raisebox{-0.15em}{BLEU(w/o Perplexity)}  & \raisebox{-0.15em}{0.751} & \raisebox{-0.15em}{0.768} & \raisebox{-0.15em}{0.805} & \raisebox{-0.15em}{0.794} \\
\raisebox{-0.15em}{Cosine(w/ Perplexity)} & \raisebox{-0.15em}{0.827} & \raisebox{-0.15em}{0.839} & \raisebox{-0.15em}{0.817} & \raisebox{-0.15em}{0.851} \\
\raisebox{-0.15em}{Cosine(w/o Perplexity)} & \raisebox{-0.15em}{0.704} & \raisebox{-0.15em}{0.693} & \raisebox{-0.15em}{0.711} & \raisebox{-0.15em}{0.767} \\
\raisebox{-0.15em}{Euclidean(w/ Perplexity)} & \raisebox{-0.15em}{0.858} & \raisebox{-0.15em}{0.852} & \raisebox{-0.15em}{0.806} & \raisebox{-0.15em}{0.842}  \\
\raisebox{-0.15em}{Euclidean(w/o Perplexity)} & \raisebox{-0.15em}{0.681} & \raisebox{-0.15em}{0.705} & \raisebox{-0.15em}{0.706} & \raisebox{-0.15em}{0.765}  \\
\raisebox{-0.15em}{Inner Product(w/ Perplexity)}  & \raisebox{-0.15em}{0.829} & \raisebox{-0.15em}{0.808} & \raisebox{-0.15em}{0.831} & \raisebox{-0.15em}{0.842} \\
\raisebox{-0.15em}{Inner Product(w/o Perplexity)} & \raisebox{-0.15em}{0.707} & \raisebox{-0.15em}{0.689} & \raisebox{-0.15em}{0.733} & \raisebox{-0.15em}{0.759} \\

\Xhline{3\arrayrulewidth}
\end{tabular}
\vspace{-5pt}
\end{table}

\begin{figure}[t]
\centering
    \begin{minipage}[t]{0.45\linewidth}
        \centering
        \includegraphics[width=\textwidth]{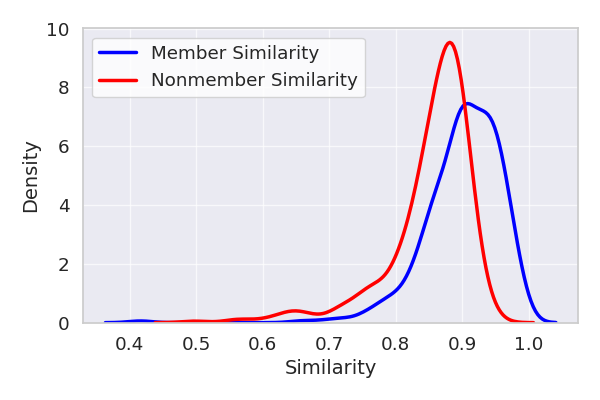}
        \centerline{\footnotesize(a) Cosine Similarity}
    \end{minipage}%
    \begin{minipage}[t]{0.45\linewidth}
        \centering
        \includegraphics[width=\textwidth]{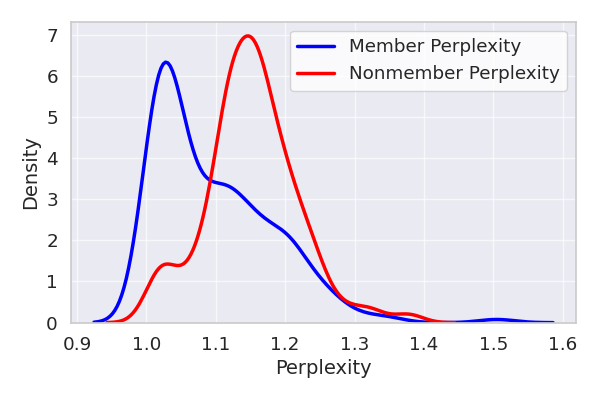}
        \centerline{\footnotesize(b) Perplexity}
    \end{minipage}
    \caption{Illustration of the distribution of member versus non-member samples}
\label{fig23}
\vspace{-5pt}
\end{figure}

\begin{table}[t]
\caption{Impact of the retriever on S$^2$MIA-T and S$^2$MIA-M in Natural Question dataset}\label{tab3}
\vspace{-5pt}
\centering
\renewcommand{\arraystretch}{0.8}
\begin{tabular}{cccccc}
\Xhline{3\arrayrulewidth}
\multirow{2}{*}{\raisebox{-0.3em}{Retriever}} & \multicolumn{2}{c}{\raisebox{-0.2em}{S$^2$MIA-T}} & \multicolumn{2}{c}{\raisebox{-0.2em}{S$^2$MIA-M}} \\
\cmidrule(r){2-3} \cmidrule(r){4-5}
 &ROC AUC & PR AUC & ROC AUC& PR AUC  \\ 
\hline
\raisebox{-0.15em}{DPR} & \raisebox{-0.15em}{0.892} & \raisebox{-0.15em}{0.872} & \raisebox{-0.15em}{0.878} & \raisebox{-0.15em}{0.882} \\
\raisebox{-0.15em}{Contreiver} & \raisebox{-0.15em}{0.857} & \raisebox{-0.15em}{0.864} & \raisebox{-0.15em}{0.870} & \raisebox{-0.15em}{0.891} \\
\Xhline{3\arrayrulewidth}
\end{tabular}
\vspace{-5pt}
\end{table}

\subsection{Results}
\subsubsection{Main Results}
We evaluate the performance of S$^2$MIA on the RAG system utilizing various LLMs, as demonstrated in Table~\ref{tab1}. The results show that S$^2$MIA can outperform all baseline methods. 
All comparisons rely on the model's training, exploiting the model's overfitting characteristics on member samples to perform MIAs. However, since RAG has not been trained on the external database, this leads to challenges for these existing methods within the RAG framework. 

\subsubsection{Impact of Different Similarity Metrics} \label{ablation}
As shown in Table~\ref{tab2}, we evaluate the effectiveness of various similarity metrics on attack efficacy. Metrics like the inner product, Euclidean distance, and cosine similarity show relatively poor performance, possibly due to the sentence embedding process which may smooth over or lose subtle semantic differences. Conversely, the BLEU score performs better. BLEU considers n-gram matching, focusing not only on individual words but also on phrases and sentence structures, providing a multi-level comparison. 
Fig.~\ref{fig23} further demonstrates that member samples exhibit higher similarity and lower perplexity compared with that of the non-members. Consequently, as detailed in Section~\ref{score}, we utilize BLEU scores and perplexity in our paper.

\begin{table}
\caption{Performance of S$^2$MIA-T and S$^2$MIA-M against Paraphrasing, Prompt Modification, and Re-ranking Defenses.}\label{tab4}
\vspace{-5pt}
\renewcommand{\arraystretch}{0.8}
\centering
\setlength{\tabcolsep}{2pt} 
\arrayrulecolor[gray]{0.7}
\begin{tabular}{c|c|cc|cc}
\arrayrulecolor{black}
\Xhline{3\arrayrulewidth}
\arrayrulecolor[gray]{0.7}
\multirow{2}{*}{\raisebox{-0.3em}{Defense}} & \multirow{2}{*}{\raisebox{-0.3em}{Metric}} & \multicolumn{2}{c|}{\raisebox{-0.2em}{Natural Question}} & \multicolumn{2}{c}{\raisebox{-0.2em}{Trivia-QA}} \\
\cmidrule(r){3-6} 
 &  & S$^2$MIA-T & S$^2$MIA-M & S$^2$MIA-T & S$^2$MIA-M \\
\midrule
\multirow{2}{*}{Paraphrasing} & \raisebox{-0.15em}{ROC AUC} & \raisebox{-0.15em}{0.563} & \raisebox{-0.15em}{0.511} & \raisebox{-0.15em}{0.606} & \raisebox{-0.15em}{0.546} \\
 & \raisebox{-0.15em}{PR AUC} & \raisebox{-0.15em}{0.504} & \raisebox{-0.15em}{0.542} & \raisebox{-0.15em}{0.575} & \raisebox{-0.15em}{0.498} \\
\arrayrulecolor[gray]{0.99}\midrule
\multirow{2}{*}{Prompt Modifying} & \raisebox{-0.15em}{ROC AUC} & \raisebox{-0.15em}{0.598} & \raisebox{-0.15em}{0.532} & \raisebox{-0.15em}{0.569} & \raisebox{-0.15em}{0.491} \\
 & \raisebox{-0.15em}{PR AUC} & \raisebox{-0.15em}{0.541} & \raisebox{-0.15em}{0.486} & \raisebox{-0.15em}{0.470} & \raisebox{-0.15em}{0.524} \\
\arrayrulecolor[gray]{0.9999}\midrule
\multirow{2}{*}{Re-ranking} & \raisebox{-0.15em}{ROC AUC} & \raisebox{-0.15em}{0.624} & \raisebox{-0.15em}{0.694} & \raisebox{-0.15em}{0.697} & \raisebox{-0.15em}{0.587} \\
 & \raisebox{-0.15em}{PR AUC} & \raisebox{-0.15em}{0.674} & \raisebox{-0.15em}{0.663} & \raisebox{-0.15em}{0.685} & \raisebox{-0.15em}{0.667} \\
\arrayrulecolor{black}
\Xhline{3\arrayrulewidth}
\end{tabular}
\vspace{-5pt}
\end{table}

\subsubsection{Impact of Different Retrievers}
In Table~\ref{tab3}, we compare the impact of DPR and Contreiver on the Natural Questions dataset, where the differences are most pronounced. Even on this dataset, which exhibits the most significant distinctions in experimental results, our S$^2$MIA demonstrates a consistently stable attack performance regardless of the RAG system's use of different configurations. 

\subsubsection{Defending against S$^2$MIA}
We assess three defense strategies: Paraphrasing~\cite{jain2023paraphrasing,yadav2024pag}, Prompt Modifying~\cite{anderson2024my,agarwal2024investigating}, and Re-ranking~\cite{zeng2024good,yu2024rankrag}. Paraphrasing rewrites the query to mislead the retriever, effectively blocking it from retrieving the original sample. Prompt Modifying changes the prompt to ``Do not directly repeat any retrieved content, but summarize it based on your understanding", which prevents the RAG from revealing the content of its external database. Re-ranking alters the order of retrieved content. 
As shown in Table~\ref{tab4}, we observe that the first two strategies effectively mitigate our attacks. In contrast, the Re-ranking strategy exhibits only a marginal decrease in attack effectiveness, suggesting that LLMs can still glean crucial information from content that has been reordered.

\section{Conclusion}
In this paper, we propose S$^2$MIA, a novel MIA against the external databases of RAG systems. Our attack does not necessitate prior knowledge of the RAG system's retriever or LLMs, nor the operation of RAG. We demonstrate that the difference in semantic similarity between member and non-member samples of RAG's external database with respect to their corresponding generated contents can compromise membership privacy. 
Experiment results show that S$^2$MIA outperforms four existing MIAs and successfully bypasses three defenses.
Our work can be viewed as a solid step in understanding the privacy leakage risks of RAG systems and shed light on designing more effective defense mechanisms against MIAs in RAG systems.

\bibliographystyle{IEEEtran}
\bibliography{references}
\end{document}